\documentstyle[12pt,epsf,aps]{revtex}  
\begin{document}
\newcommand{\beq}{\begin{equation}}
\newcommand{\eeq}{\end{equation}}
\newcommand{\beqa}{\begin{eqnarray}}
\newcommand{\eeqa}{\end{eqnarray}}
\newcommand{\Ss}{{\tilde S}}
\newcommand{\Vs}{{\tilde V}}
\newcommand{\Ts}{{\tilde T}}
\newcommand{\As}{{\tilde A}}
\newcommand{\Ps}{{\tilde P}}
\newcommand{\PVs}{{\widetilde {PV} }}
\newcommand{\fps}{f_{\pi}^2 }
\newcommand{\mks}{m_{{\mathrm K}}^2 }
\newcommand{\ms}{m_{{\mathrm K}}^{*} }
\newcommand{\msq}{m_{{\mathrm K}}^{*2} }
\newcommand{\rhos}{\rho_{\mathrm s} }
\newcommand{\rhob}{\rho_{\mathrm B} }
\newcommand{\kf}{k_{\mathrm F} }
\newcommand{\Sigs}{\Sigma_{\mathrm s} }
\newcommand{\Sigv}{\Sigma_{\mathrm v} }
\newcommand{\Sigo}{\Sigma_{\mathrm o} }
\newcommand{\mst}{ {\tilde M}^* }
\newcommand{\mstq}{ {\tilde M}^{*2} }
\newcommand{\est}{ {\tilde E}^* }
\draft
\title{Scalar and vector decomposition of the nucleon self-energy in 
the relativistic Brueckner approach} 
\author{C. Fuchs, T. Waindzoch, Amand Faessler, and D.S. Kosov}
\address{Institut f\"ur Theoretische Physik der 
Universit\"at T\"ubingen,\\ D-72076 T\"ubingen, Germany}
\maketitle  
\begin{abstract}
We investigate the momentum dependence of the nucleon self-energy in 
nuclear matter. We apply the relativistic Brueckner-Hartree-Fock 
approach 
and adopt the Bonn A potential. A strong momentum dependence 
of the scalar and vector self-energy components 
can be observed when a commonly used 
pseudo-vector choice for the covariant representation of the T-matrix
is applied. This momentum dependence is dominated by the pion 
exchange. We discuss the problems of this choice and its relations
to on-shell ambiguities of the T-matrix representation. 
Starting from a complete pseudo-vector representation of the 
T-matrix, which reproduces correctly the pseudo-vector 
pion-exchange contributions at the Hartree-Fock level, 
we observe a much weaker momentum dependence of the 
self-energy. This fixes the range of the inherent 
uncertainty in the determination of the scalar and vector 
self-energy components. 
Comparing to other work, we find that extracting the 
self-energy components by a fit to the single particle potential 
leads to even more ambiguous results. 
\\
\end{abstract}
\section{Introduction}
Concerning the nuclear many-body problem the relativistic 
Dirac-Brueckner-Hartree-Fock (DBHF) approach turned out to be 
remarkably successful in describing the nuclear matter saturation mechanism. 
The major improvement compared to non-relativistic treatments is based on an 
additional density dependence introduced in the formalism by using a  
self-consistent spinor basis. 
To solve the Bethe-Salpeter equation, i.e. its three-dimensional reductions, 
a variety of approaches relying on different techniques and various 
bare nucleon-nucleon interactions have been developed over the last decade 
\cite{horse87,thm87a,weigel88,nupp89,bm90,amorin92,boersma94,huber,sefu97,lee97}. 
All these calculations are able to describe reasonably well, although not
excellent, nuclear matter properties. In the meantime two main 
results can be regarded as settled: first, the nuclear single particle 
potential originates from the cancelation of large repulsive vector and 
attractive scalar fields, and second, the magnitude 
of the effective mass is reduced to a value of about $\sim 0.6 M$ 
at saturation density. Both findings are consistent with effective 
hadron field theories \cite{sw86} where, e.g. the effective mass 
can be determined from the spin-orbit interaction in 
finite nuclei \cite{ring}. 
Refined treatments which take into account hole-hole excitations
\cite{dejong96} or include an additional scaling 
of meson properties in the nuclear medium \cite{rapp97} do not 
significantly alter these results. 

To test the DBHF approach over a wider range of physical problems 
also finite nuclei have been studied within effective 
Hartree-Fock calculations \cite{boersma94b} and within a density dependent 
hadron field theory \cite{fuchs95,toki97}. For a successful 
application of DBHF results to heavy ion collisions 
using relativistic transport models, however, it is required 
to go beyond a local density approximation and to properly account 
for the non-equilibrium features of the highly anisotropic phase 
space configurations in such reactions \cite{fuchs96}. As soon as nuclei
overlap and particles are positioned at close quarters in 
configuration space, the momentum dependence of the nuclear 
self-energy starts to play a crucial role
in the description of heavy ion collisions.
Unfortunately, the determination of the momentum dependence of the 
nucleon self-energy is a subtle problem in the DBHF approach which has not 
yet led to settled results. Generally, the techniques applied in the 
standard relativistic Brueckner approach rely on a weak momentum dependence 
of the self-energy inside the Fermi sea. 
This assumption was supported by various calculations 
in the past \cite{horse87,thm87a,weigel88}. In Ref. 
\cite{nupp89} it was noted that the determination of the 
self-energy leads to ambiguities arising from the 
projection of the T-matrix onto  positive energy states and that the 
actual momentum dependence strongly depends on the choice of the used 
nucleon-nucleon interaction \cite{sefu97}. 
These ambiguities can be avoided when negative 
energy states are taken into account in the calculation 
\cite{weigel88,huber,fred97}. However, the conventional 
nucleon-nucleon potentials \cite{mach89} are determined 
for particle-particle scattering and an extrapolation to 
anti-particles is in itself ambiguous. 
To avoid the latter ambiguity we will 
restrict our discussion merely to the particle sector. 

In practice there exist various procedures to determine the self-energy. 
In the method proposed by Horowitz and Serot \cite{horse87} one projects onto 
Lorentz-invariant amplitudes of the T-matrix (or in-medium G-matrix) 
\cite{horse87,thm87a,nupp89,boersma94,sefu97}. This 
method is, however, not unique since pseudo-scalar (PS) and pseudo-vector (PV) 
matrix elements are equivalent for on-shell particles in the positive 
energy sector. Hence, it is impossible to
disentangle pseudo-scalar and pseudo-vector contributions in the 
on-shell T-matrix. On the other hand, the pion exchange contribution is 
known to react rather sensitively on the particular choice of its 
representation \cite{tjon85,sw86}, i.e. PS or PV. 
Thus in the past different choices for the representation 
of the corresponding amplitude have been used in the relativistic Brueckner
scheme \cite{horse87,thm87a,nupp89}.  We find that these choices strongly 
influence the magnitude and in particular the momentum dependence 
of the self-energy in the nuclear medium. In the present work we will discuss 
various possibilities to decompose the T-matrix which are based on the 
general representation of covariant amplitudes proposed in Refs. 
\cite{horse87} and  \cite{tjon85}. A strong momentum dependence of 
the self-energy is found to originate from pseudo-scalar admixtures 
due to pion-exchange contributions. These pseudo-scalar admixtures are still
present when the so called 'pseudo-vector choice' is applied, as it was done in
Refs. \cite{thm87a,nupp89,boersma94,sefu97}.
Suppressing the undesirable admixtures by making use of a 
complete pseudo-vector representation, as proposed in Ref. \cite{tjon85}, we 
find a much weaker momentum dependence of the self-energy components.  

In an altogether different approach pursued in Refs. \cite{bm90,lee97},  
the self-energy components are determined indirectly by a fit to the 
single particle potential. Here one circumvents the ambiguities of 
the projection methods. The problem is, however, only shifted to 
another level of ambiguity, since one has to extract two functions out of 
one. Thus, the fit method leads to highly ambiguous results 
concerning the momentum dependence of the self-energy components. 

The present paper is organized as follows: in section II we review
the structure of the nucleon self-energy in nuclear matter. Thereafter  
we describe in section III the representation of the T-matrix by 
Lorentz invariant amplitudes and discuss the different on-shell equivalent
choices used in the literature. 
In section IV we present new results obtained for the
different choices of the T-matrix representation utilizing the Bonn A potential
as the bare nucleon-nucleon potential. There we also discuss, as a side remark,
the approach pursued in Ref. \cite{bm90,lee97}. 
The paper ends with a summary and the conclusions of our work. 
\section{Nucleon self-energy in nuclear matter}
The properties of dressed nucleons in nuclear matter are 
expressed by the self-energy which enters  
the in-medium nucleon propagator as the formal 
solution of the Dyson equation 
\begin{equation}
 G(k) = \frac{1}{k \!\!\! / - M - \Sigma (k) +i\epsilon}
\quad .
\label{green}
\end{equation}
Due to translational and rotational invariance, parity conservation 
and time reversal invariance the self-energy in isospin saturated 
nuclear matter has the general form 
$\Sigma = \Sigma_s - \gamma_\mu \Sigma^\mu$. It depends on the Lorentz 
invariants $k^2$, $k\cdot j$ and $j^2$, with $j_\mu$ and $k_\mu$ being the
baryon current and the nucleon four-momentum, respectively \cite{sw86}. 
The invariants can also be expressed in terms of 
$k_0, |{\bf k}|$ and $\kf$, where $\kf$ denotes the Fermi momentum.
Furthermore the vector part of the self 
energy has contributions proportional to $k^\mu$ and to the current 
$j^\mu$. Defining the streaming velocity as 
$u^\mu = j^\mu / \sqrt{j^2}$, the momentum $k^\mu$ can be decomposed 
into contributions parallel and perpendicular to the streaming velocity, i.e. 
$ k^\mu = (k\cdot u) u^\mu + \Delta^{\mu\nu}k_\nu$ with the 
projector $\Delta^{\mu\nu} = g^{\mu\nu} - u^\mu u^\nu$. 
The vector part of the self-energy can then be written covariantly 
as \cite{amorin92,sehn96} 
\begin{equation}
\Sigma^\mu = \Sigo u^\mu + \Sigv  \Delta^{\mu\nu}k_\nu
\quad . 
\label{sigvec1}
\end{equation}
Thus the full self-energy reads 
\begin{eqnarray}
\Sigma (k,\kf) &=& \Sigs (k,\kf) -\gamma_\mu \left[ \Sigo (k,\kf) \, u^\mu 
+ \Sigv (k,\kf) \, \Delta^{\mu\nu}k_\nu \right]
\label{sig1} \\
&=& \Sigs (k,\kf) -\gamma_0 \, \Sigo (k,\kf) + 
{\bf \gamma}\cdot {\bf k} \,\Sigv (k,\kf)\, |_{{\mathrm RF}} 
\label{sig2}
\end{eqnarray}    
where the subscript RF indicates the respective 
expressions in the nuclear matter rest frame 
($u^{\mu} = \delta^{\mu 0}$) \cite{horse87,thm87a}. 
The $\Sigs,\Sigo$ and $\Sigv $ components are 
Lorentz scalar functions which actually 
depend on $k_0$,$|{\bf k}|$ and $\kf$. They follow from 
the self-energy matrix by taking the respective traces 
\cite{sehn96}
\begin{eqnarray}
\Sigs &=& \frac{1}{4} tr \left[ \Sigma \right] 
\label{trace1}\\
\Sigo &=& \frac{-1}{4} tr \left[ \gamma_\mu u^\mu \Sigma \right] 
       =  \frac{-1}{4} tr \left[ \gamma_0 \, \Sigma \right]_{{\mathrm RF}}
\label{trace2}\\
\Sigv &=& \frac{-1}{4\Delta^{\mu\nu}k_\mu k_\nu } 
          tr \left[\Delta^{\mu\nu}\gamma_\mu k_\nu \, \Sigma \right] 
       = \frac{-1}{4|{\bf k}|^2 } 
tr \left[{\bf \gamma}\cdot {\bf k} \, \Sigma \right]_{{\mathrm RF}} 
\label{trace3}
\quad .
\end{eqnarray}

The Dirac equation for the in-medium spinor basis can be deduced from the 
Green function (\ref{green}). Written in terms of effective 
masses and momenta 
\begin{eqnarray}
\quad M^* = M+ Re \, \Sigs \quad , \quad k^{*}_\mu = k_\mu + Re \, \Sigma_\mu 
\label{sig3}
\end{eqnarray}
the Dirac equation has the form
\begin{equation}
\left[ k^* \!\!\!\!\!\! / - M^* -i \, Im \, \Sigma \right] u(k) =0 .
\label{dirac} 
\end{equation}
In the following we will work in the quasi-particle approximation 
and therefore we neglect the imaginary part of the
self-energy from now on. Thus the effective nucleon four-momentum will be 
on mass shell even above the Fermi surface, fulfilling the relation 
$ k^{*}_\mu k^{*\mu} = M^{* 2}$. 
Since we only deal with the real part 
of the self-energy in the quasi-particle approximation we omit this in the 
notation. In the nuclear matter rest frame the four-momentum 
follows from Eq. (\ref{sig3})
\begin{eqnarray}
{\bf k}^* = {\bf k} (1+\Sigv) 
\quad , \quad
k^{*}_0 = E^* = \sqrt{ {\bf k}^2 (1+\Sigv)^2 + M^{*2} }
\label{estar}
\end{eqnarray}
which allows one to eliminate the $\Sigv$-term in the Dirac equation, 

\begin{eqnarray}
\left[ ({\bf \alpha} \cdot {\bf k}) - \gamma^0  \mst \right] u(k) =\est u(k) 
\quad ,\label{dirac2}
\end{eqnarray} 

by a rescaling of effective mass and 
energy
\begin{eqnarray} 
\mst = \frac{M^*}{1+\Sigv} \quad , \quad \est = \frac{E^*}{1+\Sigv} 
     = \sqrt{{\bf k}^2 + \mstq }
\quad . 
\label{red1}
\end{eqnarray}

From the Dirac equation in the form (\ref{dirac2})
one derives the in-medium relativistic Hamiltonian for nucleons 
and thereby the operator for the single-particle potential within
the DBHF approximation, i.e. ${\hat U } = \gamma^0  \Sigma$. 
The expectation value of ${\hat U}$, i.e. sandwiching ${\hat U}$ between 
the effective spinor basis, yields the single particle potential 
\begin{eqnarray}
   U(k) = \frac{<u(k)|\gamma^0  \Sigma | u(k)>}
{< u(k)| u(k)>} =
\frac{M^{\ast}}{E^{\ast}({\bf k})} 
\, <{\bar u(k)}| \Sigma | u(k)>
\label{upot1}
\end{eqnarray}
which can be evaluated as 
\begin{eqnarray}
U(k,\kf) &=& \frac{M^*}{E^*} \Sigs - \frac{ k_{\mu}^* \Sigma^\mu}{E^*} \\
         &=& \frac{M^* \Sigs }{\sqrt{ {\bf k}^2 (1+\Sigv)^2 + M^{*2}}} 
         - \Sigo + \frac{ (1+\Sigv)\Sigv {\bf k}^2}
           {\sqrt{ {\bf k}^2 (1+\Sigv)^2 + M^{*2}}}
\quad .
\label{upot2}
\end{eqnarray}
In many applications \cite{bm90,lee97} the single particle 
potential is only given in terms of a scalar and zero-vector component. 
This can be achieved in nuclear matter 
by introducing reduced fields ${\tilde \Sigs}$ and 
${\tilde \Sigo}$ as 
\begin{eqnarray}
{\tilde \Sigs} = \mst -M = \frac{\Sigs - \Sigv M}{1+\Sigv} 
\quad , \quad 
{\tilde \Sigo} = \est -E = \Sigo - \est ({\bf k}) \Sigv
\quad . 
\label{red2}
\end{eqnarray}
Then the single particle potential has the form
\begin{equation}
U(k,\kf) = \frac{\mst}{\est} {\tilde \Sigs} - {\tilde \Sigo} 
\quad . 
\label{upot3}
\end{equation}
Frequently the reduced fields, Eq. (\ref{red2}), are 
used rather than the projected components since they represent 
the self-energy in a mean field or Hartree form. Thus they can 
easily be related to effective hadron mean field theory 
\cite{fuchs95,toki97}. Such a representation is 
meaningful since the $\Sigv$-contribution is a moderate 
correction. 

In contrast to the single particle potential which can easily 
be derived form the T-matrix \cite{bm90} the extraction of the 
self-energy components is a subtle problem. In the latter case 
one has to represent the T-matrix within the Clifford algebra 
in the Dirac space which is not free from severe ambiguities. 
Before we discuss this point in detail we briefly recall some 
basic features of the relativistic Brueckner scheme. For more details 
see e.g. Refs. \cite{horse87,thm87a,sefu97}. 
The iteration of the Thompson equation requires to 
determine the self-consistent spinor basis, Eq. (\ref{dirac}). 
In practice the problem is treated in terms of the 
reduced effective mass $\mst$ to avoid an explicit dependence 
on the space-like $\Sigv$ components (\ref{dirac2}).
Actually, the zero-vector component $\Sigo$ does not enter 
the self-consistency problem. In the standard treatment 
\cite{horse87,thm87a,bm90,amorin92,sefu97} 
the effective mass is assumed to 
be entirely density dependent, i.e. a constant effective mass 
$\mst = \mst (|{\bf k}|=\kf,\kf)$ generally taken at the Fermi-momentum 
is used as the iteration parameter. The 
effective mass at its value at the Fermi momentum is reinserted 
into the Thompson equation and this procedure is repeated until 
convergence is reached. Such a treatment is self-consistent 
concerning the density dependence of the effective interaction 
screened by the medium and appears to 
be justified if the self-energy is weakly momentum dependent 
inside the Fermi-sea. If the self-energy is, however, strongly 
momentum dependent, as it was observed in \cite{nupp89,sefu97}, 
one principally has to go beyond the standard approach. Then one has to  
include the momentum dependence of the effective mass in the 
Thompson propagator as well as on the level of the interaction. 
For the present investigations we will not include such an explicit 
momentum dependence in the Brueckner scheme since first of all, this
is technically a rather involved problem. Secondly, we 
will verify in the next section that the momentum dependence 
of the self-energy is mainly governed by the treatment of the pion-exchange 
in the T-matrix representation. Thus a more careful treatment
of the pion-exchange leads to a much weaker momentum dependents of
the self-energy as it is desirable 
for the present self-consistency scheme. 
\section{Decomposition by Lorentz invariant amplitudes}

The self-energy for the nucleon $k$ 
follows from the T-matrix by integrating 
over the occupied states $q$ in the Fermi sea  
\begin{equation}
\Sigma (k) = -i \int \frac{d^4 q}{(2\pi )^4} 
 tr \left[ G_{{\mathrm D}}(q) {\hat T}(qk;qk - kq) \right] .
\label{sig4} 
\end{equation}
Here ${\hat T}$ denotes the T-matrix (or G-matrix) 
operator depending on four Dirac indices of the ingoing and outgoing 
nucleons. Due to antisymmetrization  ${\hat T}$ contains a 
direct ( Hartree) and an exchange (Fock) 
contribution. The Dirac propagator
\begin{equation}                                          
G_{{\mathrm D}}(q) = [q^{\ast}\!\!\!\!\! / + M^{\ast}] 2\pi i 
\delta(q^{\ast 2} - M^{\ast 2})\Theta(q^{\ast 0}) \Theta(\kf -|{\bf q}|) 
\label{diracpr}
\end{equation}
projects onto positive energy states in the Fermi sea. 
In order to project out the 
self-energy components, Eqs. (\ref{trace1}--\ref{trace3}), the 
T-matrix has to be decomposed into Lorentz invariants. Since we need to
consider only on-shell scattering of particles in eq. (\ref{sig4}), 
five invariant amplitudes with five covariants are sufficient to 
represent the on-shell T-matrix \cite{tjon85}. 
In this case, the scalar, vector, tensor, axial-vector and 
pseudo-scalar terms 
\beqa
S = 1\otimes 1 \quad , \quad 
V =  \gamma^{\mu}\otimes \gamma_{\mu}  \quad , \quad
T = \sigma^{\mu\nu}\otimes\sigma_{\mu\nu} \quad , \quad
A = \gamma_5 \gamma^{\mu}\otimes \gamma_5 \gamma_{\mu}\quad , \quad
P = \gamma_5 \otimes \gamma_5 
\eeqa 
form a linearly independent, however, not unique set of covariants. 
Using this special set, the on-shell T-matrix can be represented entirely 
by 'direct' amplitudes 
\beq
 {\hat T} (\theta) = 
  F_1 (\theta) S 
+ F_2 (\theta) V
+ F_3 (\theta) T 
+ F_4 (\theta) A  
+ F_5 (\theta) P
\quad . 
\label{cov3}
\eeq
The covariant amplitudes $F_i$ are determined from anti-symmetrized 
plane wave helicity matrix amplitudes 
\cite{mach89} which obey the selection rule $(-)^{L+S+I} =-1$. 
We solve the relativistic Thompson equation for the T-matrix 
\cite{sefu97} consistently in the two-particle center-of-mass frame. 
Since we need only diagonal matrix elements for 
the self-energy (\ref{sig4}) the $F_i$ amplitudes 
are required at the scattering angle 
$\theta = 0$. They are 
already anti-symmetrized and contain implicitly direct and 
exchange contributions. Therefore the simple representation (\ref{cov3}) 
is sufficient to calculate the self-energy. 
If we insert the T-matrix given by Eq. (\ref{cov3}) into 
Eq. (\ref{sig4}) and take the trace over the Dirac propagator 
only the scalar and vector amplitudes $F_1$ and $F_2$ survive 
and contribute to the self-energy, 
\begin{equation}
\Sigma (k) = -i \int \frac{d^3 q}{(2\pi )^3} 
\frac{ \Theta (\kf - |{\bf q}|)}{ E^* ({\bf q})}
\left[ m^* F_1 (0) + q^{\ast}\!\!\!\!\! /  \, F_2 (0) \right] .
\label{sig4b} 
\end{equation}
The representation of ${\hat T}$ is, however, not uniquely defined 
in the on-shell case and therefore various alternative possibilities
exist to construct the set of five independent covariants in the 
subspace of positive energies. Although the different 
representations discussed below are all equivalent if one works with 
the pseudo-scalar covariant $P$, their difference becomes 
crucial as soon as one switches from the pseudo-scalar to the 
pseudo-vector representation. The PV covariant in the medium is defined as
\beq
PV =  \frac{k_{2}^* \!\!\!\!\! / - k_{1}^* \!\!\!\!\! /}{2M^{\ast}}\gamma_5
\otimes 
\frac{q_{2}^* \!\!\!\!\! / - q_{1}^* \!\!\!\!\! /}{2M^{\ast}}\gamma_5
\eeq
with $k_{1}^*,q_{1}^* $ the initial and $k_{2}^*,q_{2}^* $ 
the final momenta of the scattering particles. The ambiguity 
of the decomposition procedure arises 
from the fact that the PS and PV matrix elements are identical 
(using the Dirac eq.) for on-shell 
scattering of positive energy states, i.e. 
\beq 
{\overline u}(k_2 )\left( 
\frac{k_{2}^* \!\!\!\!\! / - k_{1}^* \!\!\!\!\! /}{2M^{\ast}}  
\right) \gamma_5 u(k_1 ) = {\overline u}(k_2 )\gamma_5 u(k_1 )
\quad .
\label{mat1}
\eeq
Thus the corresponding amplitudes are identical as well and 
it is impossible to uniquely disentangle the PS and PV contributions 
in the T-matrix. 

However, it is known from meson theory of the 
nuclear interaction that a pseudo-vector representation of 
the $\pi N$ coupling is preferable \cite{mach89}. 
The influence of the pion in the 
relativistic theory has been discussed in detail, e.g. in
Refs. \cite{sw86,tjon85}. There it was shown that the one-pion exchange 
contribution to the nuclear optical potential tends to increase 
drastically at low momenta if the $\pi N$ vertex is treated as PS. 
One reason for this behavior 
is a strong PS coupling to negative energy states which is not apparent 
in non-relativistic approaches. The PV vertex suppresses the coupling 
to antiparticles since the overlap matrix elements vanish 
for on-shell scattering
\beq 
{\overline v}(k_2 )\left(   
\frac{k_{2}^* \!\!\!\!\! / - k_{1}^* \!\!\!\!\! /}{2M^{\ast}} 
\right) \gamma_5 u(k_1 ) = 0 
\quad .
\label{mat2}
\eeq
Therefore a PV vertex is more consistent with the 
approximation scheme of the conventional Brueckner model where 
the coupling to negative energy states is not taken into account. 
The PV vertex also strongly suppresses the 
pion contribution in particular at low energies which is more 
in accordance with the empirical knowledge from proton-nucleus scattering 
\cite{sw86,tjon85}. 
Consequently, the $\pi N$ vertex of the bare interaction 
is usually treated as PV \cite{mach89}. 

Due to these facts the usage of a PV covariant in the 
decomposition of the T-matrix is considered as preferable 
\cite{thm87a,sefu97}. However, simply replacing $P$ by $PV$ 
in Eq. (\ref{cov3}) leads to identical results; firstly, 
because the corresponding amplitudes are equal and secondly, 
because only the scalar and vector amplitudes $F_1$ and $F_2$ 
contribute to the self-energy (\ref{sig4b}). Motivated by this 
fact alternative representation of the T-matrix have been used 
\cite{thm87a,nupp89,sefu97} which lead to an -- in principle 
superfluous -- explicit splitting into 'direct' and 'exchange' contributions 
and give different results when the 
$PS \longmapsto PV $ replacement is performed. 

\subsection{Symmetrized representations}

Based on Ref. \cite{gold} 
Tjon and Wallace discussed a representation which accounts for 
the structure of the exchange contributions in ${\hat T}$ in 
a more transparent way \cite{tjon85}. To express exchange contributions 
also the Dirac indices of the covariants are interchanged.   
The transformation 
$\Ss$ interchanges the Dirac indices of particles 1 and 2, i.e. 
$\Ss u(1)_\sigma u(2)_\tau = u(1)_\tau u(2)_\sigma$. Thus one obtains 
the interchanged covariants as 
$\Ss = \Ss S,\Vs = \Ss V, \Ts = \Ss T, \As = \Ss A$ and $\Ps = \Ss P$.  
The interchanged covariants are related to the original covariants 
(\ref{cov3}) by a Fierz transformation ${\cal F}$ \cite{tjon85}
\beqa
\left( 
\begin{array}{c} \Ss \\ \Vs \\ \Ts \\ \As \\ \Ps \end{array} 
\right)
= \frac{1}{4}
\left( 
\small{ 
\begin{array}{ccccc} 
 1  &  1 & \frac{1}{2} & -1  &  1   \\
 4  & -2 &  0          & -2  &  -4  \\
 12 &  0 &  -2         & 0   &  12  \\
 -4 & -2 &  0          & -2  &  4   \\
 1  & -1 & \frac{1}{2} & 1   &  1 
\end{array}} 
\right)
\left( 
\begin{array}{c} S \\ V \\ T \\ A \\ P \end{array} 
\right) .
\label{fierz}
\eeqa
For a definite scattering angle $\theta$ the T-matrix is now 
represented by five symmetrized covariants \cite{tjon85}
\beq
 {\hat T} (\theta) =  
f_1 (\theta) (S - \Ss) 
+ f_2 (\theta) \frac{1}{2}(T + \Ts)
- f_3 (\theta) (A - \As)
+ f_4 (\theta) (V + \Vs) 
+ f_5 (\theta) (P - \Ps) 
\quad .
\label{cov1}
\eeq
These symmetrized covariants are constructed so that the amplitudes 
$f_i$ account for the Pauli principle in a transparent way. 
Interchanging the outgoing particles, 
anti-symmetrization requires the following symmetry 
\beq
f^{\rm I}_i (\pi-\theta) = (-)^{\rm I+ i}f^{\rm I}_i (\theta)
\label{asym}
\eeq
for definite isospin $I=0,1$. 

The relation between the five new amplitudes $f_i$ and the former amplitudes
$F_i$ is given by the transformation \cite{tjon85} 
\beqa
\left( 
\begin{array}{c} f_1 \\ f_2 \\ f_3 \\ f_4 \\ f_5 \end{array} 
\right)
= \frac{1}{4}
\left( 
\small{ 
\begin{array}{ccccc} 
 2 & -4 & -12 & 0  &  0  \\
 1 &  0 &  4  & 0  &  1  \\
 0 & -2 &  0  & -2  &  0 \\
 1 &  2 &  0  & -2 &  -1 \\
 0 &  4 & -12 & 0  &  2 
\end{array}} 
\right)
\left( 
\begin{array}{c} F_1 \\ F_2 \\ F_3 \\ F_4 \\ F_5 \end{array} 
\right) .
\label{transform}
\eeqa
With relation (\ref{asym}) one can express T as the combination of 
two terms which resemble a direct and an exchange contribution, 
i.e. ${\hat T} = {\hat T}^{\rm D} - {\hat T}^{\rm X} $, by 
\beqa
 {\hat T}^{\rm D} (\theta) &:=&  
  f_1 (\theta) S 
+ f_2 (\theta) \frac{1}{2} T
- f_3 (\theta) A 
+ f_4 (\theta) V  
+ f_5 (\theta) P  \label{cov2} 
\\
 {\hat T}^{\rm X} (\theta) &:=& (-)^{\rm I+ 1} \left[ 
  f_1 (\pi - \theta) \Ss 
+ f_2 (\pi -\theta) \frac{1}{2} \Ts
- f_3 (\pi -\theta) \As 
+ f_4 (\pi -\theta) \Vs  
+ f_5 (\pi -\theta) \Ps \right] 
\quad .
\nonumber
\eeqa
However, it can not be claimed that ${\hat T}^{\rm D}$ and 
${\hat T}^{\rm X}$ are the real direct and exchange matrix elements 
but only their combination yields the fully anti-symmetrized 
matrix elements. 

The representation proposed 
by Horowitz and Serot \cite{horse87} and also 
used in other works \cite{thm87a,sefu97} is of a similar 
structure as the one above with the difference that it is 
based on direct amplitudes 
\beqa
 {\hat T}^{\rm D} (\theta) &:=& \frac{1}{2} \left[ 
  F_1 (\theta) S 
+ F_2 (\theta) V
+ F_3 (\theta) T 
+ F_4 (\theta) A  
+ F_5 (\theta) P \right]  \label{cov4} 
\\
 {\hat T}^{\rm X} (\theta) &:=& (-)^{\rm I+ 1} \frac{1}{2}\left[ 
  F_1 (\pi -\theta) \Ss 
+ F_2 (\pi -\theta) \Vs
+ F_3 (\pi -\theta) \Ts 
+ F_4 (\pi -\theta) \As  
+ F_5 (\pi -\theta) \Ps \right]   
\quad .
\nonumber
\eeqa
To be more precise, in this approach both set of amplitudes $ F_i (\pi)$
and $ F_i (\pi -\theta)$ 
are obtained from the helicity matrix elements applying representation 
(\ref{cov3}) at angle $\theta$ and $\pi -\theta$, respectively.
Doing this, the amplitudes $ F_i $ 
are essentially different from the $f_i$ amplitudes of Eq. (\ref{cov1}). 
Instead of relation (\ref{asym}) anti-symmetrization requires now 
that the exchange amplitudes are connected to the direct amplitudes 
by the Fierz transformation (\ref{fierz}) 
\beq
F^{\rm I}_i (\pi-\theta) = (-)^{\rm I} {\cal F}_{ji} F^{\rm I}_j (\theta)
\label{asym2}
\quad .
\eeq  
The difference between the representations (\ref{cov3}) and 
(\ref{cov4}) lies in the fact that Eq. (\ref{cov4}) 
leads to an explicit splitting into direct and exchange contributions 
which, however, can be regarded 
as -- at least partially -- artificial. Since the amplitudes  
$F_i (\theta)$ and $F_i (\pi -\theta)$ in (\ref{cov4}) are 
determined from already anti-symmetrized helicity 
matrix elements one has in that case the identity 
\beq
{\hat T}^{\rm X} = - {\hat T}^{\rm D}
\quad . 
\eeq
Hence the normalization 
factor 1/2 which determines the splitting into 'direct' and 
'exchange' parts in (\ref{cov4}) is judicious and could also be 
fixed differently by any normalized linear combination. In this 
context we want to mention that concerning the original work of Horowitz and 
Serot \cite{horse87} this statement would not hold because they 
used non-antisymmetrized (unphysical) helicity states in their formalism 
and therefore they had to anti-symmetrize explicitly the T-matrix
elements by splitting the representation into direct and exchange 
contributions as done in Eq. (\ref{cov4}). As a consequence, their amplitudes
$F_i$ at angle $\pi$ and $\pi-\theta$ did not fulfill the anti-symmetry
relation (\ref{asym2}) and thus only the representation (\ref{cov4})
was physically meaningful. 

Working with physical helicity states, however,
one retains relation (\ref{asym2}) and therefore all types of 
decompositions, Eqs. (\ref{cov3}-\ref{cov4}), are equivalent as
long as one restricts to a pseudo-scalar representation. 
But if one replaces the PS by the PV covariant the 
equivalence of the expressions 
(\ref{cov3}), (\ref{cov1}) and (\ref{cov4}) is destroyed. 
Our 'optimal choice' of the representation for using 
the PV covariant will be discussed in the next section. 

\subsection{Pseudo-vector representation}
The nucleon-nucleus potential is very sensitive to the 
treatment of the pseudo-scalar or pseudo-vector treatment of the 
pion-nucleon interaction. As already mentioned, 
the pion should be preferentially treated 
with a pseudo-vector coupling in the bare 
NN interaction. The corresponding coupling strength 
is fixed in such a way that it 
reproduces the on-shell PS coupling strength in the vacuum \cite{mach89}. 
This already leads to a suppression of the vertex by 
a factor $(\frac{M^*}{M})^2 $ inside the nuclear medium \cite{sw86}. 
However, the major suppression of the pion exchange contribution 
to the nucleon self-energy originates from the different cofactors 
which arise if one inserts the Dirac propagator (\ref{diracpr}) 
and the T-matrix from (\ref{cov3}) or  (\ref{cov4}) into the 
Eq. (\ref{sig4}) for the self-energy \cite{thm87a,sefu97}, i.e. 
\beqa
tr \left[ ( q^{\ast}\!\!\!\!\! / + M^{\ast} ) \Ps  \right] 
&=& - ( q^{\ast}\!\!\!\!\! / + M^{\ast} ) 
\label{traceps}
\\ 
tr \left[ ( q^{\ast}\!\!\!\!\! / + M^{\ast} ) \PVs \right] 
&=& ( k^{\ast}\!\!\!\!\! / + M^{\ast} )
\left( \frac{ k^{*}_\mu  q^{* \mu} }{2 M^{* 2}} - \frac{1}{2} \right) 
\quad . 
\label{tracepv}
\eeqa
The influence of the pion, in particular the sensitivity on the 
PS or PV representation, is most clearly demonstrated at the 
Hartree-Fock level. This means that 
${\hat T}$ is replaced by the 
bare NN interaction ${\hat V}$ and no further medium effects are 
taken into account, i.e. bare nucleon masses are used and the 
Pauli operator in the Thompson equation is neglected. Furthermore, 
the Hartree-Fock expressions are known analytically \cite{muehf} 
which provides also a straightforward check of the involved projection 
techniques. It is a well known fact \cite{sw86} that the 
PS treatment yields extremely large pion 
contributions to the self-energy whereas the PV representation 
suppresses these terms almost completely. This effect is demonstrated 
in Fig. 1 where the Hartree-Fock contributions of the pion 
only to the nuclear self-energy are shown. The calculations are 
performed for a nuclear matter density of $\rho = 0.166 $ fm$^{-3}$ with 
a PS and PV (denoted as 'full PV' in Fig. 1) 
$\pi$NN coupling of $g^{2}_\pi / 4\pi = 14.9$ 
and the pion form factor taken from the Bonn A interaction \cite{mach89}.
The bare nucleon mass is used in this example. 
It is seen that the PS description of the 
pion exchange yields extremely large self-energy components at low 
momenta which are rapidly decreasing with increasing momentum. The 
$\Sigs$ and $\Sigo$ contributions are almost identical which leads 
to a strong cancelation effect in the single particle potential. The 
PV description (denoted as 'full PV' in Fig. 1) 
suppresses the pion by nearly two orders of magnitude 
and even on this new scale the momentum dependence is much less 
pronounced. Remarkably, scalar and vector contributions have now 
opposite signs. Thus they add up in the potential (\ref{upot3}) 
and the remaining 
momentum dependence is not the remnant of a huge cancelation effect as 
in the PS case. 

For comparison we also show the results which are 
obtained in the projection scheme when the so-called 'PV choice' 
is adopted. In the  'PV choice' commonly used \cite{thm87a,sefu97} 
we simply replace the pseudo-scalar covariants by the pseudo-vector 
covariants $P,\Ps \longmapsto PV,\PVs $ in the decomposition of 
the T-matrix (\ref{cov4}). Due to the on-shell equivalence of the respective 
matrix elements, Eq. (\ref{mat1}) the amplitudes remain thereby unchanged. 
Now the explicit choice of the T-matrix representation 
starts to play a decisive role since 
the PS and PV exchange terms contribute differently to the self-energy, 
Eqs. (\ref{traceps},\ref{tracepv}).   
Thus the strength of the PS $\longmapsto$ PV replacement,  
determined by the respective amplitudes $f_5 (\theta)$ and 
$F_5 (\pi -\theta)$, becomes important. 
In most previous calculations \cite{thm87a,sefu97} 
the representation (\ref{cov4}) was used. The effect of the
replacement can be seen in Fig. 1 where the Hartree-Fock contribution
of the pion exchange to the nucleon self-energy using the
PV choice is shown.  
First of all, it should be noted that the result using the PV choice
for (\ref{cov4}) is identical to the result which one obtains 
if one uses the replacement in the symmetrized representation, 
Eq. (\ref{cov1}), since both amplitudes $f_5 (\theta)$ and 
$F_5 (\pi-\theta)$ agree in the Hartree-Fock approach with 
only pion exchange. 
It is obviously transparent from Fig. 1 that within the 
traditional PV choice the pion is not treated correctly  
as a pseudo-vector particle. The PV choice representation 
is rather a mixture of the PS and a complete PV 
representation. Although the pion contribution to the nucleon self-energy 
is suppressed by about a factor of two compared to the original PS case, 
the structure of the self-energies, in particular the pronounced momentum 
dependence is still very similar. The reason for this behavior of the
self-energy is easily understandable. Even after replacing 
$P,\Ps$ by $PV,\PVs $, both representations, Eqs. (\ref{cov1}) 
and (\ref{cov4}), still contain pseudo-scalar admixtures because 
the Fierz transformation (\ref{fierz}) couples all covariants. 
This leads to identities for the symmetrized 
vector and tensor covariants \cite{tjon85}
\beqa
\frac{1}{2} ( T+\Ts ) &=& S+\Ss +P+\Ps 
\\    
V+\Vs &=& S+\Ss -P-\Ps 
\label{iden}
\quad . 
\eeqa
In order to completely remove the PS part from the interaction 
one first should use the identities above which leads to the 
following decomposition \cite{tjon85}
\beqa
 {\hat T} (\theta) &=& 
  ( f_1 + f_2 +f_4 ) S  
- ( f_1 - f_2 -f_4 ) \Ss  
- f_3 (A - \As)
\nonumber \\
&& 
+ ( f_5 + f_2 -f_4 ) P 
- ( f_5 - f_2 +f_4 ) \Ps 
\quad . 
\label{cov5}
\eeqa
If the replacement $P,\Ps \longmapsto PV,\PVs $ is now performed 
in Eq. (\ref{cov5}) instead of Eq. (\ref{cov1}) or Eq. (\ref{cov4})
this yields a complete PV representation of the interaction which we will 
call 'full PV' representation in Fig. 1 and in the following.
Such a decomposition can successfully describe the PV pion exchange 
on the Hartree-Fock level, i.e. the results calculated with the 
analytically known Hartree-Fock matrix elements \cite{muehf} 
for the PV pion-exchange are reproduced. 
In the present formalism this can not be achieved by the other 
decompositions. On the other hand, however, e.g. the $\omega$ exchange is 
no longer treated accurately in the full PV representation 
since PS admixtures arising from the Fock part of the $\omega$ 
exchange, Eq. (\ref{iden}), are treated as PV. This bias will, however, 
turn out to be small. Anyway, all existing 
decomposition are unable to handle the PV pion 
exchange simultaneously with the remaining set of mesons 
on the Hartree-Fock level. As we will see later on, the one-pion 
exchange dominates the momentum dependence of the self-energy. 
Hence, it is reasonable to require 
that the PV pion-exchange is treated exactly on the Hartree-Fock level. 
This constraint is fulfilled adopting the full PV treatment. 
\section{Results}

In the present section we study the impact of the various 
representations of the T-matrix (\ref{cov3}), (\ref{cov1}), (\ref{cov4}) 
and (\ref{cov5}) on the nucleon self-energy and 
on related quantities. 

\subsection{Momentum dependence of the self-energy}
On the level of the self-energy the effect of the different 
choices can be summarized as
\beqa
\Sigma (k) = \Sigma^{{\rm PS}} (k) - \delta\Sigma (k) 
= \Sigma^{{\rm PS}} (k) 
+  i \int \frac{d^4 q}{(2\pi )^4} 
f_R (kq;qk) ~tr \left[ G_{{\mathrm D}} (q) ( \Ps -\PVs ) \right]
\label{sig5}  
\eeqa
with $\Sigma^{{\rm PS}}$ being the self-energy given in the pure PS 
representation. The different approaches for the self-energy $\Sigma$ 
using for the T-matrix Eqs. (\ref{cov3}), (\ref{cov1}), (\ref{cov4}) 
or (\ref{cov5}) are only varying in the choice of $f_R$ explained 
below. The strength of the suppression of the 
pseudo-scalar contributions is moderated by the $f_R$ amplitude 
which also determines the deviation $\delta\Sigma$ of the self-energy 
from the PS case. 
In principle this deviation shift is only apparent in the 
decomposition of $\Sigma$ in the scalar and vector components, Eq. 
(\ref{sig1}), but vanishes when the complete matrix elements are built, i.e. 
\beq
<{\bar u(k)}| \delta \Sigma (k) | u(k)> = 0
\quad .
\label{iden2}
\eeq
For the same effective mass $M^* (\kf)$ all approaches give 
the same total single particle potential (\ref{upot1}) 
although they yield 
quite different scalar and vector self-energy components. 
However, the different approaches also 
yield different values for the effective mass $M^* (\kf)$ which 
leads to a different in-medium spinor basis $|u>$ used in 
the self-consistent iteration procedure and the equivalence 
for the single particle potential gets distorted. 
Consequently, the different approaches lead to visuable 
changes also for those quantities which are built from 
total matrix elements, i.e. the single particle potential and the 
equation-of-state. If the value of $M^*$ is, however, kept 
fixed the equivalence on the level of matrix elements is 
exact and also numerically fulfilled with high accuracy.    
The cases discussed in the previous section modify only $f_R$ 
in Eq. (\ref{sig5}) and can now be summarized as 
\beqa
 f_R = \left\{ 
\begin{array}{ccc}
0 & \quad , \quad & {\rm PS} \\
(-)^{\rm I+ 1} \frac{1}{2} F_5 (\pi-\theta) 
& \quad , \quad & {\rm 'PV~choice'} \\
f_5 (\theta)- f_2 (\theta) +f_4 (\theta)  
& \quad , \quad & {\rm 'full~PV'}
\end{array}
\right.
\label{choice}
\eeqa
Since earlier works on relativistic Brueckner theory which sticked 
to the projection method \cite{thm87a,nupp89,sefu97} applied 
the scheme proposed by Horowitz and Serot 
(Eq. (\ref{cov4}) with $P$ and $\Ps$ replaced by $PV$ and $\PVs$, 
respectively, our 'PV choice') we will discuss this 
method first. Within this scheme the structure of the 
self-energy, i.e. its density and momentum 
dependence has been investigated in detail in Ref. \cite{sefu97} 
with the Bonn potentials \cite{mach89} as 
the bare NN-interaction. As the most prominent result we observed a 
strong momentum dependence of the scalar and time-like 
vector self-energy components around the Fermi momentum. These findings 
are in qualitative agreement with a previous work by 
Nuppenau {\it et al.} \cite{nupp89}. 
Fig. \ref{fig2} shows the momentum dependence 
of the three self-energy components $\Sigs , \, \Sigo ,\, \kf\Sigv$ at  
nuclear matter density $\rho = 0.166\, fm^{-3}$ obtained with the 
Bonn A interaction. The space-like $\Sigv$ component is found to be 
relatively large inside the Fermi-sea and decreases
with increasing momentum. Furthermore we compare to the corresponding reduced 
fields ${\tilde \Sigs},\,{\tilde \Sigo}$ where the $\Sigv$-term is 
effectively included, Eq. (\ref{red2}). It is seen that the 
reduced fields are significantly 
smaller in magnitude at low momenta and generally show a less pronounced 
momentum dependence. The inclusion of the $\Sigv$-term 
counterbalances the strong momentum dependence to some extent. 
The negative $\Sigv$-contribution effectively weakens the 
momentum dependence whereas in the opposite 
case \cite{amorin92,fred97} 
the momentum dependence will be enhanced. In the 
limit of a vanishing $\Sigv$-contribution reduced and projected 
components are equal. 

\begin{figure}[h]
\begin{center}
\leavevmode
\epsfxsize = 21cm
\epsffile[60 110 600 590]{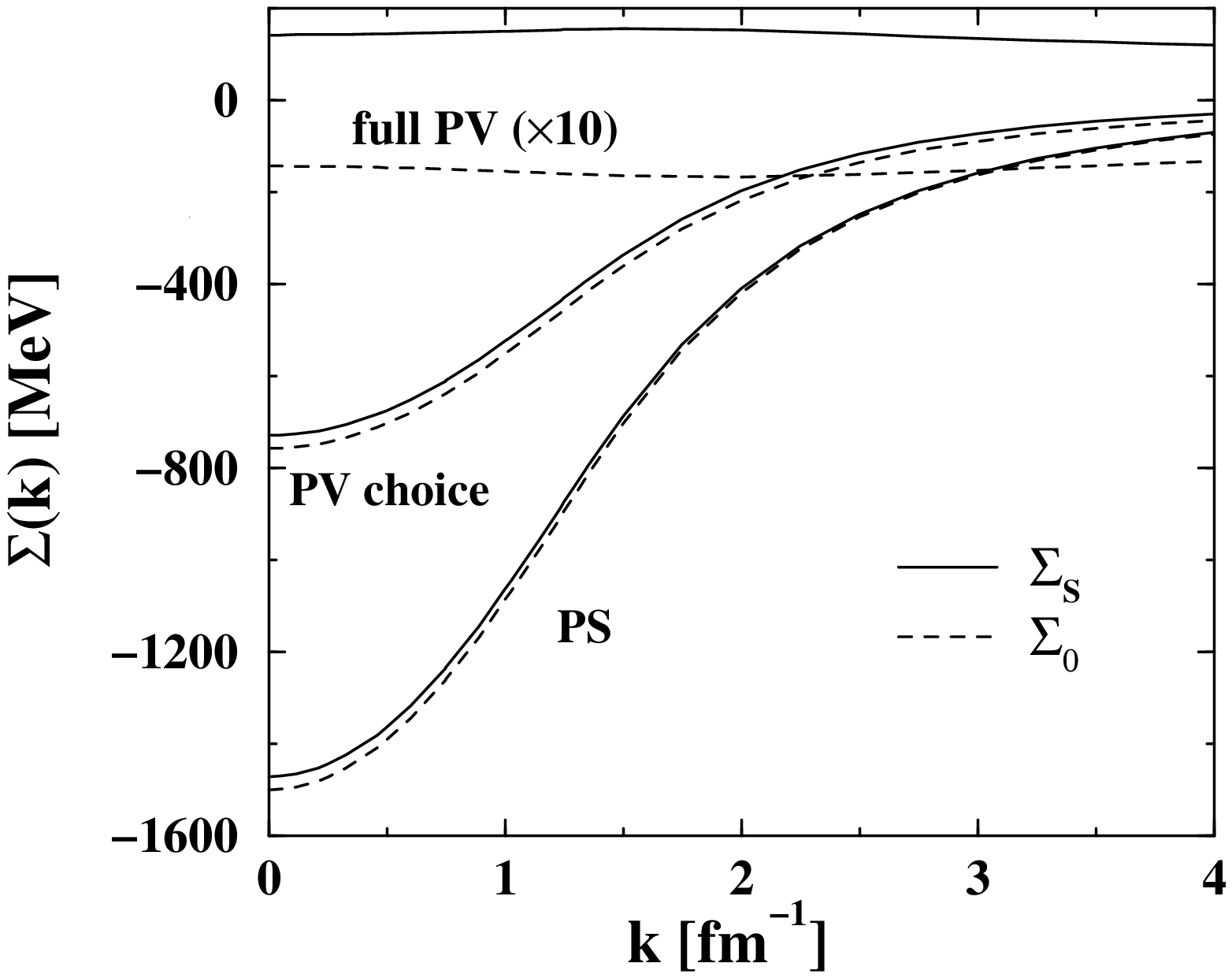}
\end{center}
\caption{\label{fig1}
}
\end{figure}

\begin{figure}[h]
\begin{center}
\leavevmode
\epsfxsize = 21cm
\epsffile[60 110 600 590]{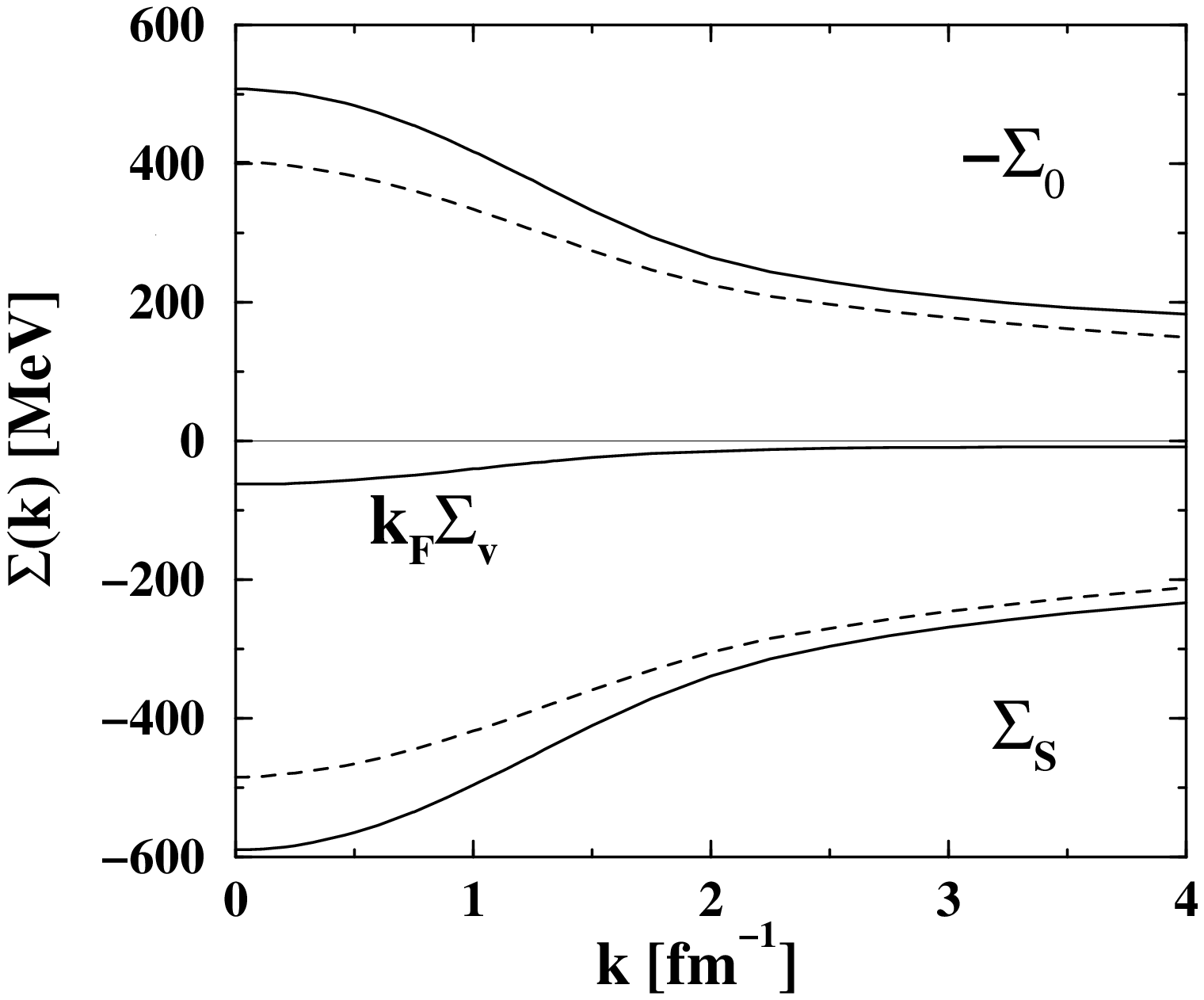}
\end{center}
\caption{\label{fig2}
}
\end{figure}

\begin{figure}[h]
\begin{center}
\leavevmode
\epsfxsize = 21cm
\epsffile[60 110 600 590]{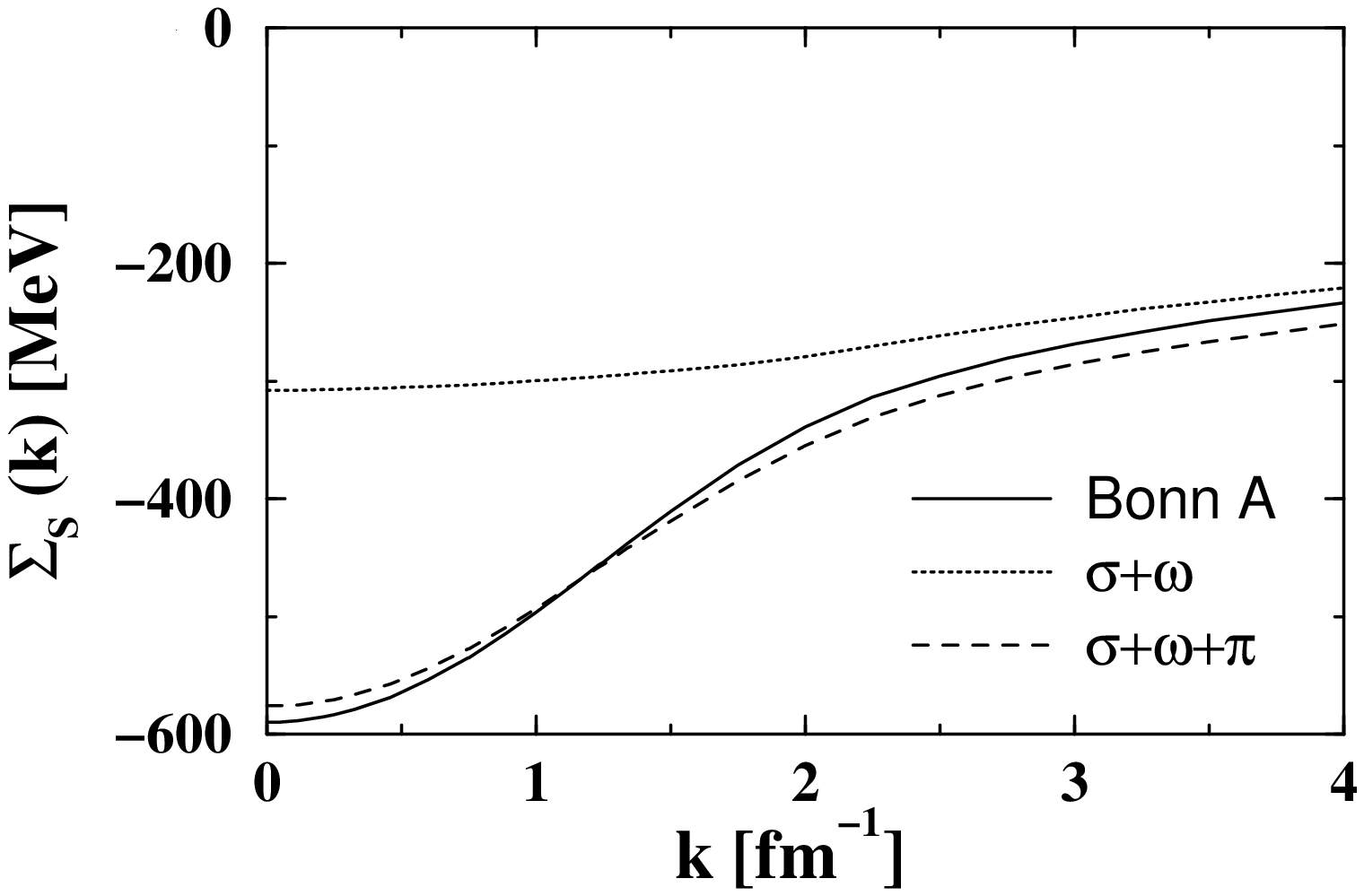}
\end{center}
\caption{\label{fig3}
}
\end{figure}

Fig.\ref{fig3} demonstrates the influence of the various mesons
(using Bonn A). 
Taking only $\sigma$ and $\omega$ exchange into account the 
result is quite similar to that obtained in Ref. \cite{horse87}, 
i.e. the momentum dependence is flat inside the Fermi sea. Including 
the pion we are already very close to the full result. The 
strong momentum dependence of the present calculation originates to 
a large extent from the pion-exchange. Hence, the calculation 
still reflects the Hartree-Fock results (Fig.\ref{fig1}) and 
the strong momentum dependence originates mainly from 
pseudo-scalar contributions of the pion. 

\begin{figure}[h]
\begin{center}
\leavevmode
\epsfxsize = 21cm
\epsffile[60 110 600 590]{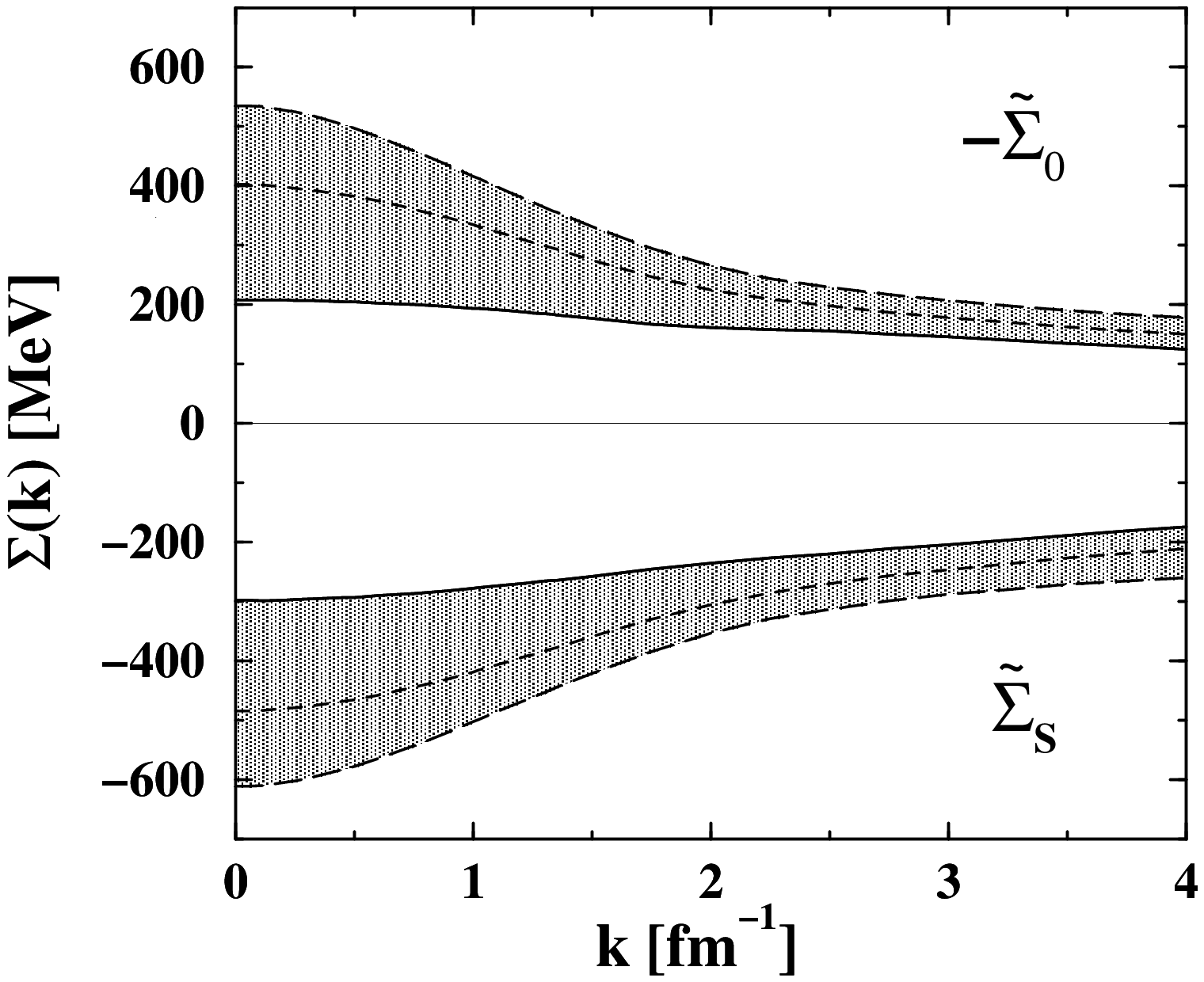}
\end{center}
\caption{\label{fig4}
}
\end{figure}

In Fig.\ref{fig4} the corresponding self-energies obtained 
for the various decompositions are compared. Adopting the 
full PV representation the space-like $\Sigv$ contribution 
turns out to be much smaller than in the PS or the 
standard PV choice (see also Fig. \ref{fig2}). 
Therefore we show the reduced self-energies 
${\tilde \Sigs}$ and  ${\tilde \Sigo}$ in which $\Sigv$ 
is included for a better comparison.  
Both, the PS and the PV choice show a similar strong 
momentum dependence which again reflects the pseudo-scalar nature of 
the pion exchange. Consistent with the previous 
considerations this momentum dependence vanishes to a large 
extent when the pion contribution is suppressed by the full PV 
representation of the T-matrix. At high energies the different 
choices yield similar results since the influence of the pion 
decreases. The pure PS and the full 
PV representation can be regarded as the limiting 
cases which give the range of uncertainty in the determination 
of the self-energy. 
The full PV representation has thereby the big advantage 
that due to the weak momentum dependence the 
standard treatment of the Thompson equation, i.e. to approximate 
the effective mass by its constant value at the Fermi surface, 
can be safely applied as done in our and almost all other works on 
this subject. Furthermore this method ensures by 
construction that it treats the pion correctly 
at the Hartree-Fock level. 
Thus the full PV representation comes closest of all discussed 
representations to what we would call the optimal representation
of the T-matrix. It is also worthwhile to mention that the 
results obtained within this representation agree well with 
recent calculations which include negative energy 
states and thus avoid the 
projection procedure \cite{fred97}.
\subsection{Single particle potential and the fit method}

As discussed above, the single particle potential, Eq. (\ref{upot1}), 
is in principle independent on the representation of the T-matrix which 
is again due to the on-shell equivalence of the corresponding 
matrix elements, Eq. (\ref{mat1}). 
However, significantly different values 
of $M^*$ obtained in the different approaches lead  
to different results. Although this effect is 
reduced using the rescaled effective 
mass $\mst $, the equation-of-state 
reacts sensitively on this $M^*$ dependence, as can be seen from 
table 1. A suppression of the PS contributions causes a larger effective 
mass and thus suppresses relativistic effects which 
originate from the mixing of small and large components of the spinors. 
A pure PS treatment leads to more binding 
and shifts the saturation point to higher densities. The corresponding 
equation-of-state is rather stiff. By the empirical knowledge of 
the nuclear saturation properties the PS representation can be ruled out 
since the saturation density is much too high and the effective mass 
is much too small. The remaining two methods are in rough agreement 
with the empirical constraints, however, the densities are also 
slightly too high compared to the experimental Fermi momentum 
of about $\kf = 1.37 fm^{-1}$. 
Here the larger value of the effective mass 
is a favor of the 'full PV' representation since it is in better 
agreement with the constraints derived from the spin-orbit 
splitting in finite nuclei \cite{ring}. Compared to the 'PV choice', 
the 'full PV' representation leads to more binding and makes the 
equation-of-state softer, in agreement with the incompressibility 
derived from the isoscalar giant monopole resonance of about 
K $\approx$ 230 MeV. One might assume that this is due to the 
fact that the 'full PV' representation suppresses part of the repulsive 
$\omega$-exchange. This is however, not the case since the inaccuracy 
which arises in the $PS \longmapsto PV$ replacement 
procedure concerning the vector and 
tensor exchange amplitudes, Eqs. (\ref{iden}), has only a minor influence 
on the final results. We checked this point by treating the 
Hartree-Fock contribution of the one-pion-exchange separately in the 
PV representation while for the remaining part of the T-matrix 
the PS representation was retained. In terms of the self-energy this 
means to set $f_R = f^{X}_\pi $ in (\ref{sig5}) with $f^{X}_\pi $ 
the Hartree-Fock amplitude of the one-pion exchange.

In view of the problems which arise in the determination of 
the self-energy we now briefly discuss a frequently used and much simpler 
approach first applied by Brockmann and Machleidt 
\cite{bm90}. In this approach one tries to extract the self-energy components 
directly from the single particle potential,  
thus one avoids to take the explicit traces (\ref{trace1}--\ref{trace3}).
Therefore there is no need to decompose the T-matrix into its 
Lorentz invariants. 
Actually, Brockmann and Machleidt determined 
only density dependent but not momentum 
dependent mean 
values for ${\tilde \Sigs},\, {\tilde \Sigo}$ by a fit to $U$  
given by Eq. (\ref{upot3}). This 
fit method works reasonably well as long as one restricts oneself to 
density dependent observables on the one-body level. 
The corresponding (reduced) effective  
masses are relatively close to our results at $\kf$ 
obtained within the 'PV choice' \cite{sefu97}. Thus it is 
understandable that the resulting nuclear matter saturation 
properties are similar in the two approaches of 
Refs. \cite{bm90} and \cite{sefu97}, 
see Tab.1. However, any information on the 
magnitude of the space-like $\Sigv$-contribution and the 
explicit momentum dependence of the self-energy is completely lost 
in this approach. To overcome this drawback 
the Stony Brook group \cite{lee97} extracted 
momentum dependent fields from the single particle potential. 
They assumed a functional 
dependence of the (reduced) self-energies of the form
\begin{equation}
{\tilde \Sigma_{{\mathrm s,o}}} (k) = 
\frac{ \alpha_{{\mathrm s,o}} }{ 1+\beta_{{\mathrm s,o}} 
(k/\kf)^{\gamma_{{\mathrm s,o}}} }
\label{fit1}
\end{equation}
at fixed density $\kf$. The set of six parameters 
$\{\alpha ,\beta , \gamma \}_{{\mathrm s,o}}$ was then determined by 
a least square fit to $U$, i.e. by minimizing 
\begin{equation}
 \chi^2 = \int_{0}^{\kf} dk k^2 \left[ U(k,\kf) - 
\left(\frac{\mst}{\est} {\tilde \Sigs} - {\tilde \Sigo} \right)
\right]^2
\quad . 
\label{fit2}
\end{equation}
However, such a procedure suffers from a large amount of 
arbitrariness since 
one tries to extract two independent functions out of one function. 
Consequently, the result is strongly influenced by the choice of the 
trial functions. It is clear that the class of trial functions 
given by Eq. (\ref{fit1}) will generally not provide solutions 
of Eq. (\ref{sig5}), although the insertion of the -- a priori 
unknown -- 'correct' amplitude $f_R$ should yield the correct 
results for the self-energies. 

\begin{figure}[h]
\begin{center}
\leavevmode
\epsfxsize = 21cm
\epsffile[60 110 600 590]{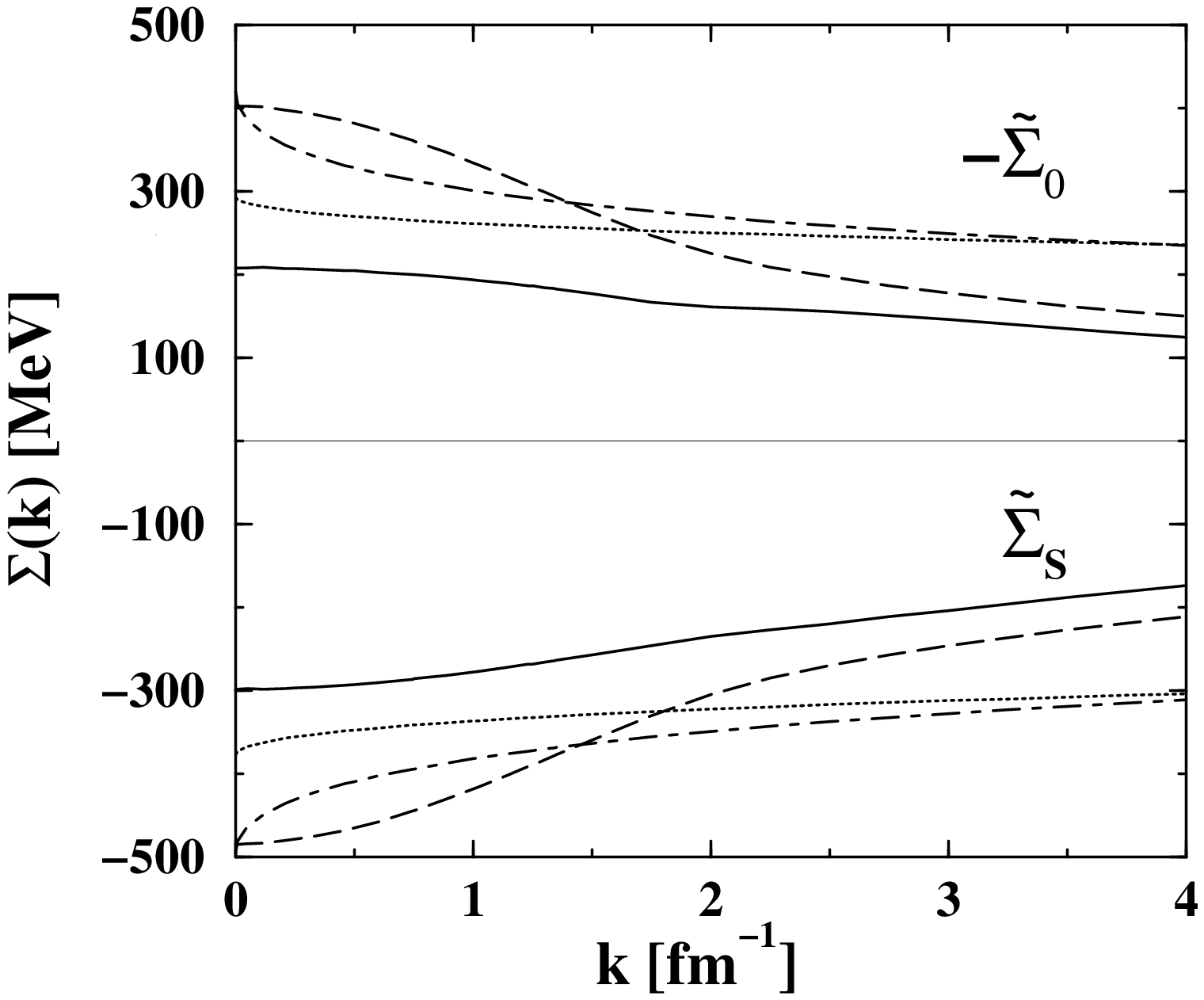}
\end{center}
\caption{\label{fig5}
}
\end{figure}

To demonstrate 
this aspect we compare in Fig. \ref{fig5} the fit procedure  
according Eqs. (\ref{fit1},\ref{fit2}) to the projection 
method using thereby both, the 'PV choice' and the 'full 
PV' decomposition. The fitted self-energies are determined from the 
single particle potential obtained with the 'PV choice'. 
It is seen that the fitted self-energies 
show a moderate momentum dependence which is in a 
qualitative agreement with the 'full PV' representation, 
but not with the 'PV choice' to which the fit was performed.  
However, the asymptotic high momentum behavior of 
the fitted self-energies 
is completely different from the projected self-energies, 
independently which choice is used.  
Also the slope of the curves at 
low momenta seem to be distorted by the choice of the trial 
function. The results of Ref. \cite{lee97} show a 
similar behavior (which is due to 
the choice of the same functional $k$-dependence (\ref{fit1})), 
however, the momentum dependence 
is a little more pronounced than our fitted results. 
For a fair comparison one has to be 
aware that in Ref. \cite{lee97} and 
the present work different approximations to the Thompson 
propagator were made. Actually, in 
the present work the Thompson equation is solved in the 
two-particle center-of-mass frame, whereas in Refs. 
\cite{bm90,lee97} it is solved in the nuclear matter rest frame 
thus avoiding the projection techniques. The main difference 
is, however, that in Ref. \cite{lee97} 
the effective mass entering into the Thompson equation 
and the effective spinor basis is allowed to be 
momentum dependent. This states an involved problem which 
affords a couple of additional approximations. If the momentum 
dependence of the self-energy is very pronounced one has  
to go beyond the present approximation scheme with $\mst (\kf)$ 
in order  
to include this momentum dependence self-consistently. However, 
performing the full PV decomposition (which we regard as the 
most reliable one) the momentum dependence is actually very weak 
(see Fig. 4). Thus the usage of a constant effective mass 
$\mst (|{\bf k}|=\kf,\kf)$ is well justified. 
In contrast to \cite{lee97} where it was argued that the improved 
self-consistency suppresses the momentum dependence we find that 
momentum dependence is mainly governed by the representation of 
the T-matrix or -- in the case of Ref. \cite{lee97} -- by the 
choice of the trial functions.   

\begin{figure}[h]
\begin{center}
\leavevmode
\epsfxsize = 18cm
\epsffile[110 110 500 500]{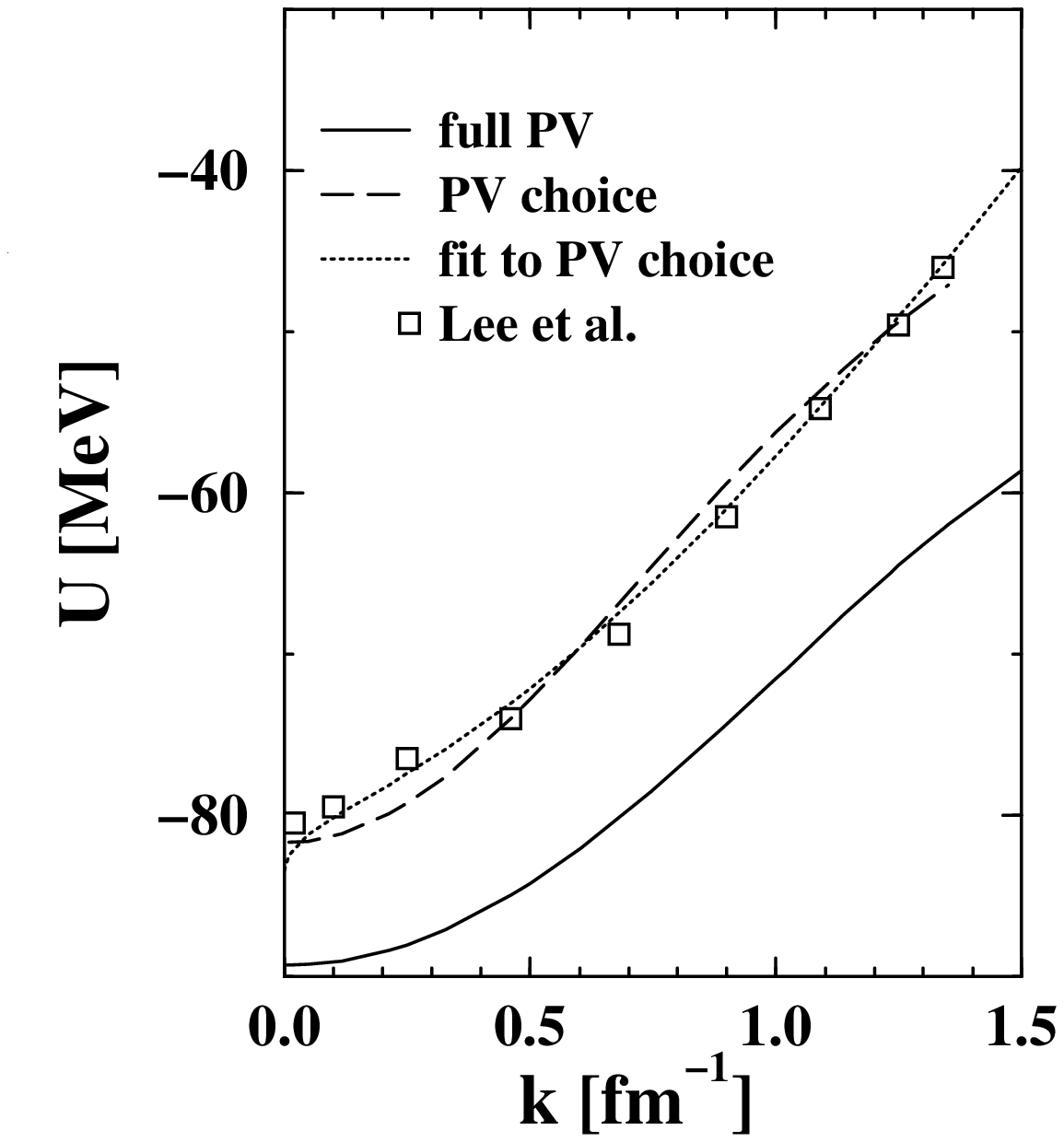}
\end{center}
\caption{\label{fig6}
}
\end{figure}

To illustrate this effect the single particle 
potential is considered in Fig. \ref{fig6}. 
It is seen that the different decompositions, 
'PV choice' versus 'full PV' representation, yield  
significantly different results for this quantity. The deviations 
are due to the different modifications of the effective interaction
in the self-consistency scheme which yield also different 
effective masses. The 'full PV' treatment 
lowers the potential by about 8 MeV 
compared to the standard treatment and thus leads to 
more binding in the equation-of-state. We also 
included the result of Ref. \cite{lee97} into this figure. 
Remarkably, this calculation yields the same result as 
the present 'PV choice' 
although the underlying scalar and vector 
self-energies are completely different. 
This somehow fortuitous agreement can be understood by the fact 
that the self-energies coincide around the Fermi momentum. 
In addition we show in Fig. \ref{fig6} 
the single particle potential obtained as a result of the 
fit procedure, Eq. (\ref{fit2}). $U$ is reasonably well 
reproduced although the fitted and 
projected self-energies in Fig. 5 strongly differ. Hence, 
it is not possible to extract the self-energy components 
from the single-particle potential in a reliable way. 
This finding is also supported by a recent analysis 
where it was shown that the fit method breaks down when applied to 
isospin asymmetric nuclear matter \cite{ulrych97}. Recently 
M\"uther, Ulrych and Toki \cite{muether98} also determined 
momentum dependent scalar and vector self-energies components 
directly from the single particle potential (\ref{upot3}). There 
$\mst $ was treated as an independent quantity 
which allows to generate more than one equation for $U(k,\kf)$ 
to determine ${\tilde \Sigs}$ and 
${\tilde \Sigo}$. This approach, however, neglects that only 
the self-consistent $\mst$ has a physical meaning. With the 
method of \cite{muether98} one also obtains a very weak momentum 
dependence of ${\tilde \Sigs}$ and ${\tilde \Sigo}$ 
similar to the 'full PV' approach of 
the present work. 

Another experimentally accessible observable is the Schroedinger equivalent 
optical potential \cite{thm87a}. Here the explicit momentum 
dependence of the self-energies provides an important correction 
to the in first order linear energy dependence of the optical 
potential. Adopting the 'PV choice' we already found 
in Ref. \cite{sefu97} a good agreement with 
the empirical values of Ref. \cite{hama90} for the real part 
up to energies around 800 MeV and an excellent agreement 
up to the pion threshold for the imaginary part. As can be seen from Fig. 7 
where the real part of the optical potential is shown the 
agreement with the data is even improved at low energies when 
the 'full PV' representation is used. This also indicates that 
the depth of the corresponding single particle potential 
is quite reasonable. At higher energies the two methods yield 
almost identical results. On the other hand, with the 
fields as obtained by the fit method and 
also predicted similar in \cite{lee97} it is 
not possible to reproduce the high energy behavior of the 
optical potential. This will of course have implication when 
such fields, Eq. (\ref{fit1}), are applied to the 
transport description of heavy ion collisions. 

\section{Summary and Conclusions}
In the present work we investigated the momentum dependence of the 
nuclear self-energy in the relativistic Brueckner approach. We applied 
the standard treatment which projects onto positive energy states and 
determines the self-energy components by a decomposition of the 
T-matrix into its Lorentz invariants. The T-matrix is 
represented by a set of five linearly independent covariants. 
Since the set of covariants is not uniquely determined in the 
subspace of positive energies one has some 
freedom in the choice of the representation. It is therefore not 
possible to determine the scalar and vector parts of the 
relativistic nuclear self-energy in a unique way. This ambiguity 
originates from the fact that the on-shell scattering of positive 
energy states yields identical values for the 
pseudo-scalar and the pseudo-vector 
representation of the corresponding matrix elements and that 
they connect differently to the negative energy states, 
see also Refs. \cite{huber,fred97}. 
In the standard treatment of relativistic Brueckner theory 
one accounts for this fact by 
choosing a particular type of a pseudo-vector representation. 
To perform this 'PV choice', see Eq. (\ref{cov4}),  
one has to decompose already 
anti-symmetrized matrix elements into direct and exchange 
terms which is not free from ambiguities. 
Applying the Bonn potentials as the bare NN interaction 
we find that the conventional 
'PV choice' leads to a pronounced momentum dependence of 
the nuclear self-energy. The spatial $\Sigv$-contribution 
of the self-energy is thereby found to be relatively large, 
in particular inside the Fermi sea, and counterbalances 
this momentum dependence to some extent on the mean field 
level. However, this strong momentum dependence 
makes the standard Brueckner approach questionable 
since the Thompson equation (or alternative reductions) are iterated 
using a self-consistent effective mass depending only on the Fermi 
momentum. 

The momentum dependence is found to be 
completely dominated by the pion exchange. 
It originates from the pseudo-scalar nature of the pion 
which is still remnant adopting the 'PV choice' (\ref{cov4}) in the 
conventional manner. To eliminate this insufficiency we 
represented the T-matrix by a pure pseudo-vector decomposition and 
called this 'full PV' representation. 
The 'full PV' representation accounts 
correctly for the desired pseudo-vector nature of the pion exchange 
and suppresses the momentum dependence of the self-energy 
almost completely. The two limiting cases, namely the full 
pseudo-scalar and the full pseudo-vector representation, set the 
range of uncertainty in the determination of the self-energy. 
However, the 'full PV' representation is more consistent with the 
usual approximation scheme of the Brueckner approach which 
assumes that the screening of the effective interaction in the 
medium introduces an additional density dependence, but is only 
weakly depending on the momentum of the particles. 
We further investigated a frequently 
used method, namely to determine the scalar and vector 
self-energy components directly by a fit to the single 
particle potential. Although this method works reasonably well as long 
as one restricts to density dependent observables it leads to highly 
ambiguous results when applied to extract the full momentum dependence 
of the fields. Thus we conclude that the projection onto covariant 
amplitudes using thereby a complete pseudo-vector representation 
is up to now the most reliable way to determine the scalar and 
vector nucleon self-energy components as long as one works 
exclusively with positive energy states. 
 
\begin{acknowledgments}
We would like to thank H. M\"uther, E. Schiller and H. Lenske for 
enlighting discussions. We further thank E. Schiller and H. M\"uther 
for providing us with a relativistic Hartree-Fock 
program which was very useful 
in order to check the present results. 
\end{acknowledgments}

\newpage
\begin{table}
\caption{
Binding energy per particle $E$, Fermi momentum $\kf$, reduced effective 
mass $\mst$, and compression modulus $K$ for nuclear matter 
at saturation density employing the various 
representations of the T-matrix (\ref{choice}). 
As the bare NN interaction the 
Bonn A potential was used.  
}
\begin{tabular}{ccccc}
      & $\kf$       &   E    &  $\mst$ &   K   \\
      & [fm$^{-1}$] & [MeV]  & [MeV]   & [MeV] \\
\tableline
 PS         & 1.45        & -17.70  &  455.6    & 335 \\
 PV choice  & 1.41        & -15.81  &  538.3    & 275 \\
 full PV    & 1.42        & -16.59  &  648.8    & 245 \\
 Ref. \cite{bm90} & 1.40  & -15.59  &  564.3    & 290 \\
\end{tabular}
\label{tab1}
\end{table}

\section*{Figure Captions}

\begin{itemize}
\item[Fig.\,1:] Pion contributions only to the nucleon 
self-energy in the Hartree-Fock approximation. 
The self-energies determined in the pure 
pseudo-scalar (PS) or the pure pseudo-vector 
(PV, multiplied by a factor 10) representation are 
compared to results obtained within the 'PV choice'. 
In all calculations the nuclear matter 
density is chosen as $\rho=0.166\,{\mathrm fm}^{-3}$ and 
the pion coupling constant and form factor from 
the Bonn A are used. Solid lines represent the scalar, 
dashed lines the vector self-energy. 

\item[Fig.\,2:] Momentum dependence of the nucleon self-energy 
in the relativistic Brueckner-Hartree-Fock approach using 
the 'PV choice' (see (\ref{choice}). The nuclear matter 
density is chosen as $\rho=0.166\,{\mathrm fm}^{-3}$ and 
the Bonn A potential is used. Solid lines represent the scalar, 
zero-vector and space-like vector components $\Sigs ,\,\Sigo$ and 
$\kf\Sigv$; dashed lines represent the corresponding reduced fields into 
which the $\Sigv$-term is effectively included (\ref{red2}).

\item[Fig.\,3:]
Influence of the various meson exchange contributions on the 
nucleon self-energy (scalar part). The solid line corresponds to 
the full calculation (Bonn A), the dotted line to $\sigma\omega$ and 
the dashed line to $\sigma\omega\pi$--exchange only. In all 
calculations the 'PV choice' is used. 

\item[Fig.\,4:]
Comparison of the reduced self-energy components, 
Eq. (\protect\ref{red2}), obtained 
by the various decompositions of the T-matrix (see (\ref{choice}). 
Solid lines 
correspond to the 'full PV ' representation, dashed lines to 
the 'PV choice' and long-dashed lines to a pure 
PS representation. The shaded area indicates the range of 
uncertainty spanned by the 'full PV' and the PS representation, 
Eq. (\ref{cov3}). 
The nuclear matter density is chosen as 
$\rho=0.166\,{\mathrm fm}^{-3}$ and the Bonn A potential is used.

\item[Fig.\,5:]
The reduced self-energy components obtained 
with the projection method using the 'PV choice' 
(long-dashed, see (\ref{choice})) 
are compared to the respective 
fields obtained by a fit to single particle potential 
(\ref{fit1}) (dotted). 
In addition the results with the 'full PV' representation (solid)  
and those of Ref. \protect\cite{lee97} (dot-dashed) 
are shown. The nuclear matter density is chosen as 
$\rho=0.166\,{\mathrm fm}^{-3}$ and the Bonn A potential is used. 

\item[Fig.\,6:]
Single particle potential in the relativistic Brueckner-Hartree-Fock 
approach with the Bonn A NN interaction at $\rho=0.166\,{\mathrm fm}^{-3}$. 
The results of the projection method with the 'full PV' representation 
(solid line, see (\ref{choice})) 
and using the 'PV choice' (dashed line) are shown. 
The dotted line represents the single 
particle potential obtained by the fit procedure 
(\ref{fit1}) and (\ref{fit2}) to the result 
of the 'PV choice'. For comparison 
also the calculation of Ref. \protect\cite{lee97} is shown. 

\item[Fig.\,7:]
Schroedinger equivalent optical potential for Bonn A at 
$\rho=0.166\,{\mathrm fm}^{-3}$. The results of 
the projection method with the 'full PV' representation 
(solid line, see (\ref{choice})) 
and using the 'PV choice' (dashed line) are shown. 
The dotted line represents the result 
obtained with the fit procedure 
(\ref{fit1}) and (\ref{fit2}) to $U$. 
The empirical values (diamonds) are 
taken from \protect\cite{hama90}. 

\end{itemize}
\newpage

\begin{figure}[h]
\begin{center}
\leavevmode
\epsfxsize = 18cm
\epsffile[110 110 500 500]{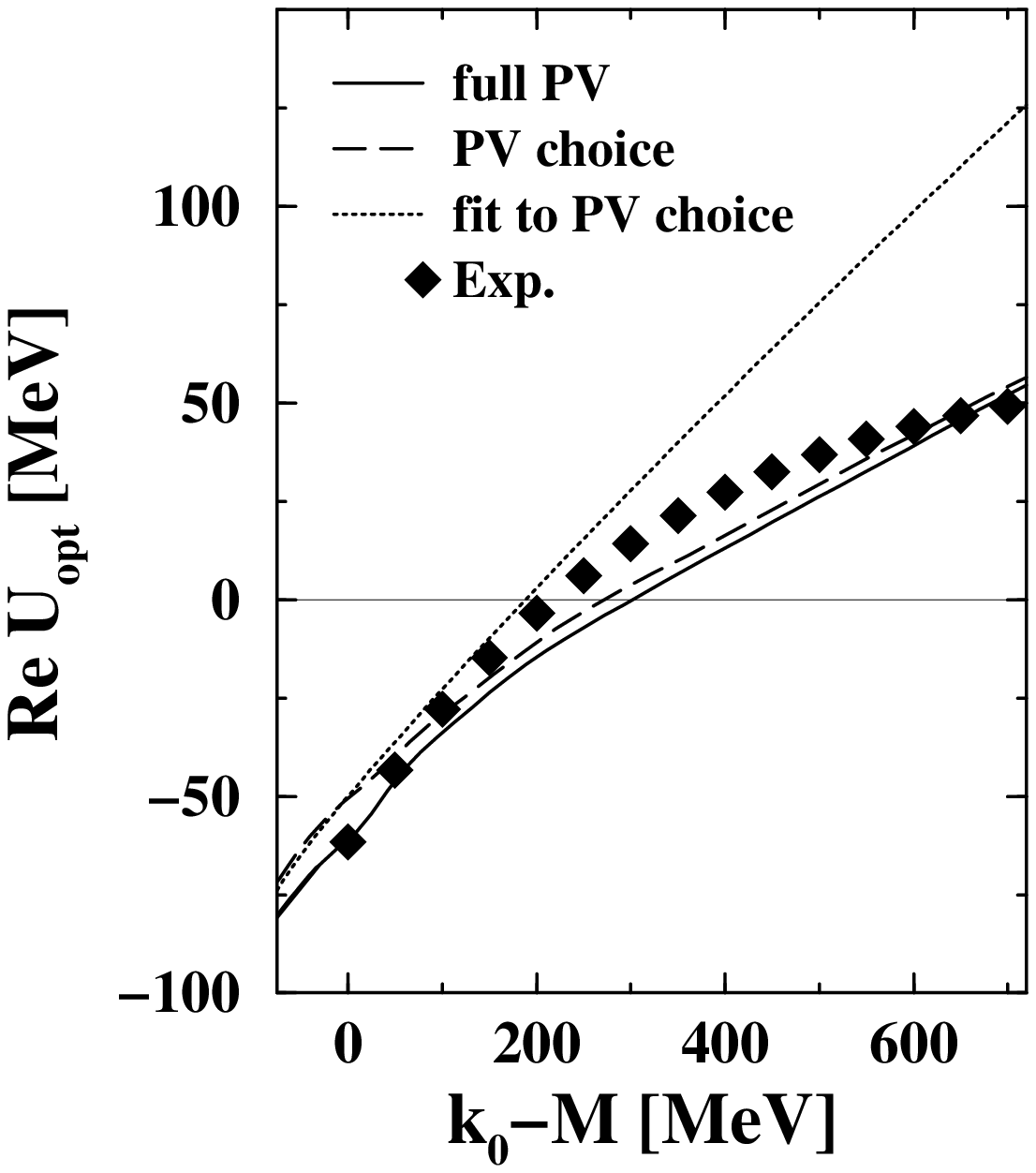}
\end{center}
\caption{\label{fig7}
}
\end{figure}


\begin{thebibliography}{99}

        \bibitem{horse87}
	C.J. Horowitz, B.D. Serot, Nucl. Phys. {\bf A464}, 613 (1987).

	\bibitem{thm87a}
	B. ter Haar, R. Malfliet, 
	Phys. Rep. {\bf 149}, 207 (1987).


	\bibitem{weigel88}
        P. Poschenrieder, M.K. Weigel,
        Phys. Rev. C {\bf 38}, 471 (1988).

        \bibitem{nupp89}
        C. Nuppenau, Y.J. Lee, A.D. MacKellar,
        Nucl. Phys. {\bf A504}, 839 (1989).  

        \bibitem{bm90}
        R. Brockmann, R. Machleidt, 
        Phys. Rev. C {\bf 42}, 1965  (1990).

        \bibitem{amorin92}
        A. Amorin, J.A. Tjon, Phys. Rev. Lett. {\bf 68}, 772  (1992).

	\bibitem{boersma94}
	H.F. Boersma, R. Malfliet, 
	Phys. Rev. C {\bf 49}, 233 (1994).

        \bibitem{huber}
	H. Huber, F. Weber, and M.K. Weigel, 
        Nucl. Phys. {\bf A596}, 684 (1995). 

        \bibitem{sefu97}
         L. Sehn, C. Fuchs, Amand Faessler,
        Phys. Rev. C {\bf 56}, 216 (1997). 

        \bibitem{lee97}
        C.-H. Lee, T.S. Kuo, G.Q. Li, and G.E. Brown, 
        Phys. Lett. B {\bf 412} (1997) 235. 

       	\bibitem{sw86}
	B.D. Serot, J.D. Walecka, 
	Advances in  Nuclear Physics, {\bf 16}, 1,
        eds. J.W. Negele, E. Vogt, (Plenum, N.Y., 1986)

        \bibitem{ring}
        P. Ring, Prog. Part. Nucl. Phys. {\bf 37}, 137 (1996).

        \bibitem{dejong96}
        F. de Jong, H. Lenske, Phys. Rev. C {\bf 54}, 1488  (1996).

        \bibitem{rapp97}
        R. Rapp, R. Machleidt, J.W. Durso, and G.E. Brown, nucl-th/9706006.

	\bibitem{boersma94b}
	H.F. Boersma, R. Malfliet, 
	Phys. Rev. C {\bf 49}, 1495 (1994).

	\bibitem{fuchs95}
        H. Lenske, C. Fuchs,
        Phys. Lett. B {\bf 345}, 355 (1995);\\
        C. Fuchs, H. Lenske, H.H. Wolter,
        Phys. Rev. C {\bf 52}, 3043 (1995).

	\bibitem{toki97}
        H. Shen, Y. Sugahara, and H. Toki,
        Phys. Rev. C {\bf 55}, 1211 (1997).

        \bibitem{fuchs96}
        C. Fuchs, T. Gaitanos, H. H. Wolter,
        Phys. Lett. B {\bf 381}, 23 (1996).

        \bibitem{fred97}
        F. de Jong and H. Lenske, preprint nucl-th/9709012. 

        \bibitem{mach89}
	R. Machleidt, 
	Advances in  Nuclear Physics,  {\bf 19}, 189,
        eds. J.W. Negele, E. Vogt, (Plenum, N.Y., 1989).

        \bibitem{tjon85}
        J.A. Tjon, S.J. Wallace,
        Phys. Rev. C {\bf 32} (1985) 267.

        \bibitem{sehn96}
	L. Sehn, H.H. Wolter, Nucl. Phys. {\bf A601}, 473 (1996);\\        
	C. Fuchs, L. Sehn, H.H. Wolter, 
        Nucl. Phys. {\bf A601}, 505 (1996). 

        \bibitem{gold}
        M.L. Goldberger, M.T. Grisaru, S.W. Mac Dowell, and D. Wong, 
        Phys. Rev. {\bf 120} (1960) 2250.

        \bibitem{muehf}
        R. Fritz and H. M\"uther, Phys. Rev. C {\bf 49} (1994) 633. 

        \bibitem{hama90}
        S. Hama, B.C. Clark, E.D. Cooper, H.S. Sherif, R.L. Mercer,
        Phys. Rev. C {\bf 41} (1990) 2737.

        \bibitem{ulrych97}
        S. Ulrych, H. M\"uther, Phys. Rev. C {\bf 56} (1997) 1788. 

        \bibitem{muether98}
        H. M\"uther, S. Ulrych, H. Toki, nucl-th/9711010. 

	\end{thebibliography}
\end{document}